\documentclass[12pt,preprint,aps]{revtex4}
\usepackage[T2A]{fontenc}
\usepackage{amssymb,amsfonts,amsmath}
\usepackage{graphicx}
\usepackage{color}
\usepackage{epsfig}
\usepackage{subfigure}
\usepackage{bm}
\usepackage{epstopdf}

\newcommand{\nix}[1]{}
\usepackage{color}

\def\be{\begin{equation}}
\def\ee{\end{equation}}
\def\bea{\begin{eqnarray}}
\def\eea{\end{eqnarray}}

\begin{document}
\title{Exciton spectroscopy of semiconductors by the method of optical harmonics generation}

\author{D. R. Yakovlev$^{1,2}$, V. V. Pavlov$^2$, A. V. Rodina$^2$, R. V. Pisarev$^2$, \\
J. Mund$^1$, W. Warkentin$^1$, and M. Bayer$^{1,2}$}
\affiliation{$^1$Experimentelle Physik 2, Technische Universit\"{a}t Dortmund, 44227 Dortmund, Germany}
\affiliation{$^2$Ioffe Institute, Russian Academy of Sciences, 194021 St. Petersburg, Russia}

\begin{abstract}
Nonlinear optical phenomena are widely used for the study of semiconductor materials. The paper presents an overview of experimental and theoretical studies of excitons by the method of optical second and third harmonics generation in various bulk semiconductors (GaAs, CdTe, ZnSe, ZnO, Cu$_2$O, (Cd,Mn)Te, EuTe, EuSe), and low-dimensional heterostructures ZnSe/BeTe. Particular attention is paid to the role of external electric and magnetic fields that modify the exciton states and induce new mechanisms of optical harmonics generation. Microscopic mechanisms of harmonics generation based on the Stark effect, the spin and orbital Zeeman effects, and on the magneto-Stark effect specific for excitons moving in an external magnetic field are considered. This approach makes it possible to study the properties of excitons and to obtain new information on their energy and spin structure that is not available when the excitons are investigated by linear optical spectroscopy. As a result of these studies, a large amount of information was obtained, which allows us to conclude on the establishing of a new field of research --- exciton spectroscopy by the method of optical harmonics generation.

This work was supported by the Deutsche Forschungsgemeinschaft (grant ICRC TRR160, project C8 and grant TRR142, projects B01 and B04) and the Russian Foundation for Basic Research (grants 15-52-12015 and 16-02-00377).
\end{abstract}
\maketitle\thispagestyle{empty}

\newpage
\setcounter{page}{1}

\section {Introduction}

The fundamental concept of an exciton as a non-current electron Bose excitation was introduced by Ya.~I.~Frenkel~\cite{Frenkel1, Frenkel2}.
Excitons of large radius, known as the Wannier-Mott excitons~\cite{Wannier, Mott}, were discovered experimentally in the semiconductor Cu$_2$O by E.~F.~Gross and H.~A.~Karryev~\cite{Gross52}. In the State Register of Discoveries of the USSR, a new discovery under the number 105, ''Exciton in semiconductors and dielectrics'',   was established with a priority of two dates: 1931 --- prediction of the existence of an exciton, and 1951 --- experimental observation of the exciton~\cite{reestr}. Excitons determine the optical properties of semiconductors near the upper edge of the forbidden band, and therefore the understanding of their properties is of great importance both for the fundamental and applied research~\cite{Gross62, Knox, Reynolds, Rashba, Yu, Klingshirn}. In quantum-size semiconductor structures, the role of excitons is even more important because the overlap of the electron and hole wave functions increases with decreasing dimensionality. In turn, this leads to an increase in the energy of their Coulomb interaction, i.e. to an increase in the binding energy of the exciton. In addition, because of the lowering of the dimensionality, the symmetry of the exciton states also changes, which leads to a strong adjustment of their energy and spin structures~\cite{Ivchenko97, Ivchenko05, Spin_book}.

Optical methods were and remain the main ones in the study of excitons. Such methods of linear optics as absorption, transmission, reflection, photoluminescence, Faraday and Kerr effects, ellipsometry are widely used. The methods of nonlinear optics are two-photon absorption, optical harmonics generation, two-photon photoluminescence excitation, generation of high-density excitons and exciton-polariton condensate, four-wave mixing, etc.~\cite{Shah, Klingshirn, Froehlich94, Bredihin, Moskalenko, Kavokin} are used to study exciton states. The use of external perturbations such as a mechanical stress, electric and magnetic fields, allows obtaining new information about excitons.

It should be noted that in the apparent similarity of the phenomena of two-photon absorption and second harmonic generation (SHG), when in both cases two photons excite an exciton, these methods are related to the different optical processes. Two-photon absorption excites all the exciton transitions allowed in the electric-dipole (ED) approximation, while the SHG should be allowed simultaneously both for the two-photon absorption process and for the one-photon emission process. It is important to note that SHG is a coherent process in which the absorption of two photons at the fundamental frequency and the emission of one photon at a doubled frequency occurs without loss of coherence of the exciton state, i.e. coherence times are usually in the range from hundreds of femtoseconds to picoseconds.

The simplest nonlinear optical processes are the second harmonic generation and third harmonic generation (THG), which convert the fundamental frequency $\omega$ of the laser into coherent radiation emitted by the crystal at frequencies $2\omega $ and $3\omega$, respectively. The optical harmonics generation in solids is widely used in laser technology~\cite{Shen, Boyd}. In the application to semiconductors, SHG and THG allow obtaining a rich information about the electronic structure in a wide spectral range~\cite{Sipe, Sipe1}. In the ED approximation, SHG is symmetry allowed only in crystals without an inversion center, while THG is possible both in centrosymmetric and noncentrosymmetric crystals. It should be noted that for quite a long time after the discovery of optical harmonics generation~\cite{Franken, Maker}, experimental studies of these phenomena in semiconductors were mostly limited to measurements on fixed laser wavelengths, when the energies of the pump photons and the generated radiation fell into the transparency region.

It was demonstrated on several groups of magnetic dielectrics that the method of the optical second harmonic generation is a powerful tool for studying the electronic and magnetic structures of crystals~\cite{Fiebig}. However, systematic spectroscopic studies of excitons in semiconductors by the method of optical harmonics generation actually began in our works since the year 2004. In these studies, laser was tuned at a frequency $\omega$ in the transparency region in such a way that the frequencies $2\omega$ and $3\omega$ of optical harmonics were in resonance with the exciton states, that was accompanied by a resonance enhancement of the generated radiation. Several groups of bulk semiconductors were studied, such as diamagnetic GaAs, CdTe, ZnO, Cu$_2$O, diluted magnetic (Cd,Mn)Te, magnetic EuTe and EuSe, including the first study of quasi-two-dimensional excitons in quantum-well ZnSe/BeTe structures~\cite{Pavlov05PRL,Pavlov05JOSAB,Saenger06a,Saenger06PRL,Saenger06b,Kaminski09,Kaminski10,Pisarev10,
Lafrentz10,Lafrentz12,Lafrentz13PRL,Lafrentz13,Brunne15,Yakovlev15}. It was shown that the modification of the exciton states in the external electric and/or magnetic fields makes them active for SHG, and can substantially enhance their contribution to THG. New mechanisms of optical harmonics generation, specific for excitons, have been disclosed and a microscopic consideration of these mechanisms has been carried out.
This article provides a brief overview of these studies.

\section{Phenomenological description of the optical harmonics generation}
\label{sec:phenomenology}

In semiconductors with a noncentrosymmetric crystal lattice, such as GaAs (the crystallographic point group $\overline{4}3m$) and ZnO (the crystallographic point group $6mm$), SHG process is allowed in the ED approximation. The nonlinear polarization at the doubled frequency $2\omega$ for the crystallographic contribution to SHG, $\mathbf{P}^{2\omega}$, can be represented as
\begin{equation}
P_{i}^{2\omega}= \epsilon_0 \chi^{\mathrm{cryst}}_{ijl}E_j^\omega E_l^\omega  ,
\label{eq:P1}
\end{equation}
where $i, j, l$ are Cartesian indices, $\epsilon_0$ is the vacuum permittivity, $\chi^{\mathrm{cryst}}_{ijl}$ is the nonlinear optical susceptibility, $E^{\omega}_{j(l)}$ is the component of the electric field $\mathbf{E}^{\omega}$ of the laser beam at the fundamental frequency $\omega$.
Equation~(\ref{eq:P1}) takes into account only the ED resonant and nonresonant contributions of the electronic states of a semiconductor at the basic $\omega$ and 2$\omega$ frequencies. A more general approach can take into account magnetic dipole (MD) and/or electric quadrupole (EQ) contributions. These contributions become important when SHG at a frequency of 2$\omega$ is in resonance with the energy of the exciton state ${\cal E}_{\mathrm{exc}}$. To account for these contributions, the effective nonlinear polarization at the doubled frequency 2$\omega$, appearing under the influence of the electric field of the electromagnetic wave at the fundamental frequency $\mathbf{E}^\omega(\mathbf{r},t) = \mathbf{E}^\omega
\exp[{\mathrm{i}(\mathbf{k}^{\omega}\mathbf{r} -\omega t)}]$ can be written in the form
\begin{eqnarray}
P_{\mathrm{eff},i}^{2\omega}({\cal E}_{\mathrm{exc}})=\epsilon_0
\chi^{\mathrm{cryst}}_{ijl}({\cal
E}_{\mathrm{exc}},\mathbf{k}_{\mathrm{exc}}) E_j^\omega E_l^\omega,
\label{eq:P2}
\end{eqnarray}
where the nonlinear susceptibility $\chi^{\mathrm{cryst}}_{ijl}({\cal E}_{\mathrm{exc}}, \mathbf{k}_{\mathrm{exc}})$ takes into account effects of spatial dispersion in the MD and EQ approximations. $\mathbf{k}_{\mathrm{exc}}=2n\mathbf{k}^{\omega}$ is the exciton wave vector, $n$ is the refractive index of light at the fundamental frequency $\omega$, and $\mathbf{k}^{\omega}$ is the wave vector of the incoming light.

The symmetry of the exciton states changes under the influence of external electric fields ($\mathbf{E}$) and magnetic fields ($\mathbf{B}$) that can mix different types of exciton states and induce new types of SHG mechanisms. Taking into account external fields, the nonlinear effective polarization at the doubled frequency $2\omega$ can be written as follows
\begin{equation}
P_{\mathrm{eff,B,E},i}^{2\omega}({\cal E}_{\mathrm{exc}})=
\epsilon_0 \chi_{ijl}({\cal
E}_{\mathrm{exc}},\mathbf{k}_{\mathrm{exc}},\mathbf{B},\mathbf{E})
E_j^\omega E_l^\omega.
\label{eq:P3}
\end{equation}
In this equation, the nonlinear susceptibility $\chi_{ijl}({\cal E}_{\mathrm{exc}},
\mathbf{k}_{\mathrm{exc}},\mathbf{B},\mathbf{E})$ takes into account the contributions of external fields. In some cases considered in Section~\ref{sec:electric}, it is more convenient to use another form of this equation in which the contributions from the scalar value of the external field and from its direction~\cite{Brunne15} are separated.
Nonlinear polarization from equations~(\ref{eq:P1})-(\ref{eq:P3}) leads to a SHG signal with the intensity $I^{2\omega}\propto |\mathbf{P}^{2\omega}|^2$.

In the case of a resonant contribution to SHG, which includes optical transitions between the ground state of the unexcited crystal $|G\rangle$ and the exciton state $|\mathrm{Exc}\rangle$, this process must be allowed both for two-photon excitation and for one-photon emission of SHG. The fulfillment of this condition depends on the symmetry of the crystal and on the geometry of the experiment. The involvement of excitons makes this picture more complicated and interesting due to the different symmetry of the envelopes of the wave functions of the $s$, $p$ and $d$ exciton states, which supplement the symmetry given by the point group of the crystal lattice. In addition, the application of uniaxial mechanical stress, electric or magnetic fields, can lead to mixing of exciton states and a lowering in their symmetry. Such effects in the experimental study of SHG on excitons can be used as tools that can change the symmetry and density of exciton states for possible resonant amplification of the nonlinear signal of SHG.

Optical harmonics generation is possible under the conditions of energy and momentum conservation. For the SHG they are written as: ${\cal E}^{2\omega}=2{\cal E}^{\omega}$ and $\mathbf{k}^{2\omega}=2\mathbf{k}^{\omega}$. In crystals where the phase matching condition $n^{2\omega}=n^{\omega}$ is satisfied, the laser and harmonic beams have the same phase velocity, which leads to an amplification of the SHG intensity. Most commercial second-harmonic generators are built on the effect of phase synchronism. In Section~\ref{sec:ZnSe_polariton} it will be shown that an anomalous dispersion in the region of exciton-polariton states makes it possible to satisfy the phase-matching condition in crystals, while this condition is not satisfied for the non-resonant SHG.

A similar phenomenological approach can be applied to the THG process. In this case, the expression for effective nonlinear polarization can be written in the form
\begin{equation}
P_{\mathrm{eff,B,E},i}^{3\omega}({\cal E}_{\mathrm{exc}})=
\epsilon_0 \chi_{ijlk}({\cal
E}_{\mathrm{exc}},\mathbf{k}_{\mathrm{exc}},\mathbf{B},\mathbf{E})
E_j^\omega E_l^\omega  E_k^\omega.
\label{eq:P33}
\end{equation}
This nonlinear polarization leads to a THG signal with the intensity $I^{3 \omega}\propto |\mathbf{P}^{3\omega}|^2$. The conditions for the conservation of energy and momentum for THG are: ${\cal E}^{3\omega}=3{\cal E}^{\omega}$ and $\mathbf{k}^{3\omega}=3\mathbf{k}^{\omega}$.  The phase matching condition $n^{3\omega}=n^{\omega}$, as a rule, can not be satisfied for the most optically uniaxial crystals. Nevertheless, in Section~\ref{sec:THG_GaAs} it will be shown that the phase matching condition for THG can be satisfied in GaAs in a magnetic field in the region of anomalous dispersion of exciton-polariton states.

\section{Experimental measurement of the optical second and third harmonics generation}
\label{sec:experiment}

The experimental setups for studying the SHG and THG spectra do not differ significantly from each other~\cite{Saenger06a, Lafrentz13, Lafrentz10, Brunne15}. The main part of the experimental setup is a pulsed laser system. To excite nonlinear processes, a large peak intensity of laser pulses is important. In this case, short pulses of about 100~fs duration, generated by femtosecond lasers with a standard repetition rate of about 80~MHz, are quite optimal. However, the use of laser amplifiers with a reduced repetition rate down to 1-100~kHz allows one to increase the peak intensity of laser pulses by several orders of magnitude while maintaining the average radiation power. The maximum pulse energy up to several tens of millijoules is achieved in lasers operating at a frequency of 10~Hz and having a pulse duration of about 8~ns. The peak intensity of nanosecond laser pulses is smaller than that for femtosecond laser amplifiers with a reduced repetition frequency. The advantage of nanosecond systems is the small spectral width of laser generation <1~meV and the possibility of computer-controlled scanning the wavelength of radiation in a wide spectral range of 0.4-2.5~$\mu$m. This is very convenient for spectroscopic studies of exciton states in various semiconductors with band gaps in range of 1-4~eV. However, a low repetition rate of the 10 Hz laser requires long signal accumulation times, and a strong laser exposure can heat the sample. Laser pulses of 100~fs duration with a high repetition frequency allow one to reduce the accumulation time and ensure a gentle laser exposure. The large spectral width of laser generation 10-20~meV, at the first glance, does not allow to achieve the required high spectral resolution of narrow exciton resonances. It will be shown that this limitation is not essential when using a spectrometer and a multichannel detector for registration of SHG and THG signals~\cite{Yakovlev15}.

The laser beam excites the SHG and THG signals in the sample, which are analyzed by a photodetector --- photomultiplier, photodiode, or CCD (charge-coupled device) camera attached to a spectrometer. The spectral dependencies are measured by scanning the wavelength of a nanosecond laser, while by using the spectrometer it is possible to filter out weak signals of SHG or THG from luminescence excited by two-photon or three-photon  absorption. When the sample is excited by the spectrally broad pulses of a femtosecond laser, it is possible to obtain the SHG and THG spectra either by scanning a spectrometer and using a single-channel detector or by recording the part of the spectrum using a CCD camera attached to a spectrometer without an exit slit. We note, that optical harmonics detection can be performed both in the transmission and reflection geometries~\cite{Saenger06a}.

Measurements of SHG and THG on exciton states in semiconductors can be carried out over a wide temperature range from liquid helium to room temperature~\cite{Saenger06a, Lafrentz13}, but for detailed spectroscopy of excitons low temperatures $T = $1.6-10~K are required. In our experiments, the sample was placed in a helium cryostat with a superconducting magnet, which made it possible to apply magnetic fields up to 10~T both in the Faraday geometry (the field is parallel to the wave vector of light) and in the Vogt geometry (perpendicular orientation). The external electric field was applied through the contacts perpendicular to the propagation direction of the light $\mathbf{E} \perp \mathbf{k}^{\omega}$. To measure the rotational anisotropies of SHG and THG (the dependences of intensity of SHG and THG signals as a function of light polarization azimuth), linear polarizers were used in the excitation and detection channels which were rotated about the direction of the light wave vector. As a rule, the signals were detected in two configurations, with rotation of either parallel ($\mathbf{E}^{\omega} \parallel \mathbf{E}^{2\omega} $) or crossed ($\mathbf{E}^{\omega} \perp \mathbf{E}^{2\omega} $) linear polarizers. Rotational anisotropies provide an important information for identifying the mechanisms of SHG and THG, as well as the corresponding exciton states~\cite {Lafrentz13, Brunne15}.

\begin{figure}[h!]
    \centering
   \includegraphics[width=10cm]{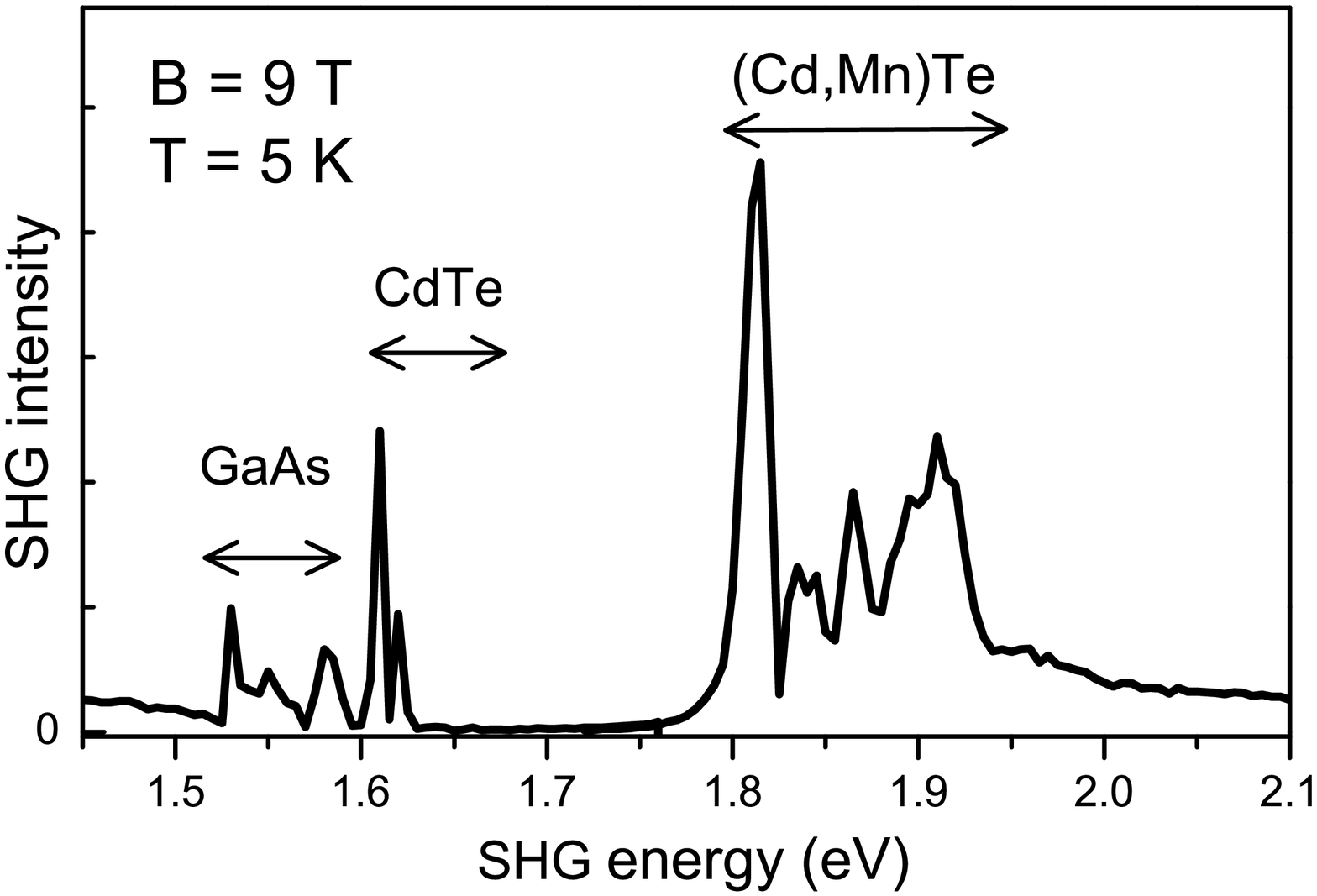}
    \caption{The spectrum of the SHG signal for a heterostructure grown by molecular-beam epitaxy on a (001)-substrate of GaAs ($E_g = 1.52$~eV), including a CdTe layer (thickness 1~$\mu$m, $E_g = 1.61$~eV) and  Cd$_{0.85}$Mn$_{0.15}$Te layer (thickness 1~$\mu$m, $E_g = 1.81$~eV). The spectrum is obtained by scanning the photon energy for laser pulses of  8~ns duration and a repetition rate of 10~Hz. A magnetic field of 9~T is applied in the Voigt geometry $\mathbf{B} \perp \mathbf{k}^{\omega} \parallel [001]$. In this geometry, the SHG signal vanishes for $B = 0$. The magnetic-field-induced SHG spectrum at $B=9$~T is shown. The arrows indicate signals for different layers of the heterostructure.}
    \label{fig:Fig1_SHG_signal}
\end{figure}

An example of the SHG spectrum from a heterostructure consisting of CdTe layers (the width of the band gap is $ E_g = 1.61 $~eV) and Cd$_{0.85}$Mn$_{0.15}$Te layer ($E_g = 1.81$ ~ eV) grown on (001)-oriented GaAs substrate ($E_g = 1.52$~eV) is shown in Fig.~\ref{fig:Fig1_SHG_signal}. In zero magnetic field, SHG for $\mathbf{k}^{\omega}\parallel [001]$ is forbidden and no signal is detected; however, a SHG signal induced by an external magnetic field is observed. The scanning energy range of laser photons 0.7-1.1~eV falls in the transparency band of all materials of this heterostructure, which allows one to measure SHG in all layers and in a substrate in a single experiment. In this experiment, SHG was recorded from the side of the wide-gap  Cd$_{0.85}$Mn$_{0.15}$Te layer in order to avoid signal reabsorption.

\begin{figure*}[h!]
    \centering
    \includegraphics[width=12cm]{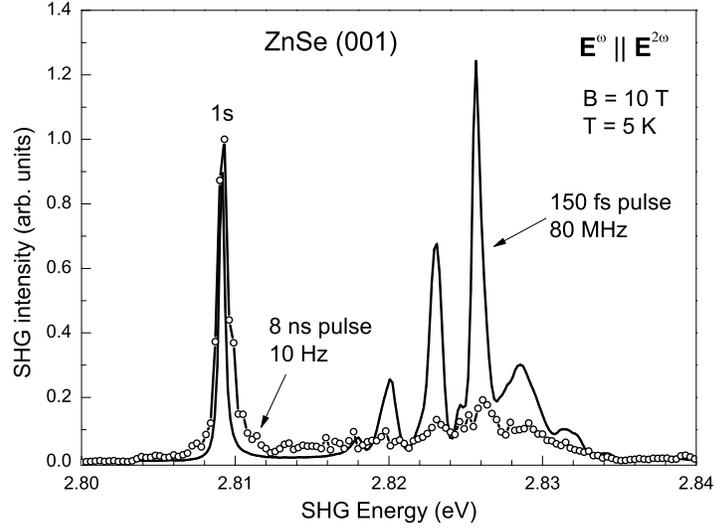}
    \caption{A comparison of the SHG spectra for magneto-exciton states in ZnSe (001), measured in two regimes: (1) photon energy scanning for laser pulses of 8~ns duration and a repetition frequency of 10~Hz, the results are represented by dots, the measurement time of the total spectrum was 1~hour; (2) the excitation of SHG signals was carried out by spectrally broad laser pulses of 150~fs duration and a repetition rate of 80~MHz with a maximum at photon energy of 1.413~eV; the spectrum is shown by a line, the exposure time was 5~min on a CCD camera combined with 50~cm spectrometer. The magnetic field is applied in the Voigt geometry  $\mathbf{B} \perp \mathbf{k}^{\omega} \parallel [001]$. The resonance at the ground state of the exciton-polariton $1s$ is at the photon energy of 2.809~eV. Resonances of magneto-excitons are seen at higher energies in the range of 2.82-2.83~eV.}
    \label{fig:Fig2_ZnSe_fs_vs_ns}
\end{figure*}

A comparison of two types of measurement of SHG spectra induced by an excitation of nanosecond and femtosecond laser pulses is shown in Fig.~\ref{fig:Fig2_ZnSe_fs_vs_ns} for the magnetic-field-induced SHG in bulk ZnSe. Despite the large spectral width of femtosecond pulses of about 15~meV, the use of a spectrometer and a CCD camera allowed us to resolve narrow peaks of magneto-exciton states. The half-width of the 1s-exciton line is  0.5~meV only. The accumulation time of the SHG spectrum in this experiment was 5 min. A good signal-to-noise ratio was obtained, which is due to  the high repetition rate of the laser of 80~MHz. In this experiment, the maximum of the laser radiation was fixed for the photon energy 1.413~eV. Using spectrally narrow nanosecond laser pulses of high power (8~ns, 10~Hz), the laser beam was scanned in the range of 1.40-1.42~eV, and the measurement time of the SHG spectrum was 1~hour. However, in the nanosecond experiments neither spectral resolution nor an improved signal-to-noise ratio were obtained.

\section{Experimental results on the optical second and third harmonics generation on exciton resonances}
\label{sec:resonance}

In this Section, experimental spectra of SHG and THG at exciton resonances in various semiconductors will be discussed. Particular attention will be paid to optical harmonics induced by external electric and magnetic fields. In order to distinguish induced contributions from crystallographic SHG signals one can take an advantage of the fact that even in the noncentrosymmetric crystals, for example, in GaAs and ZnSe, it is sufficient to choose a high-symmetry axis $[001]$ along which the crystallographic SHG signal is forbidden in the ED approximation. In this case, it is possible to establish unambiguously the appropriate mechanisms for SHG conrtibutions induced by external fields acting on the exciton states and modifying the symmetry of their wave functions.

\subsection{Optical second and third harmonics generation induced by an electric field on the $1s$-exciton in GaAs}
\label{sec:electric}

\begin{figure}[h!]
    \centering
    \includegraphics[width=16cm]{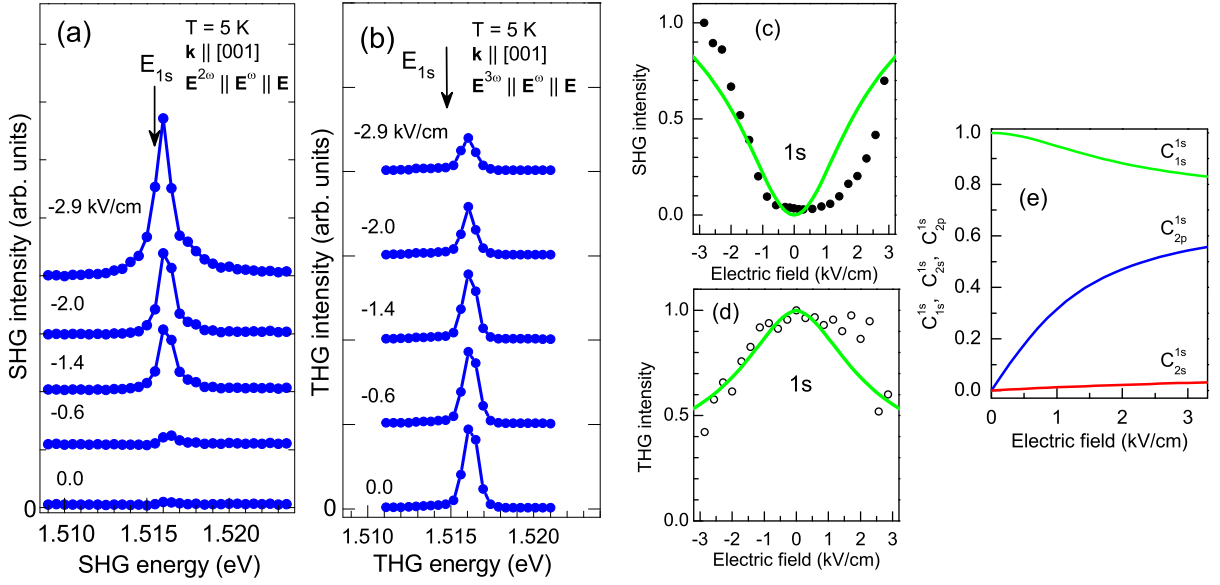}
    \caption{Effect of the electric field applied in the geometry $\mathbf{E} \perp \mathbf{k}^{\omega} \parallel [001]$ on SHG and THG in GaAs on the $1s$-exciton. (a,b) SHG and THG spectra for various values of the applied electric field obtained by scanning the photon energy for laser pulses of 8~ns duration and a repetition rate of 10~Hz. (c), (d) Dependencies of the SHG and THG signal intensities on the electric field. Symbols correspond to the experimental data and  lines are calculated using Eqs.~(\ref{eq:susceptibility2a}) and (\ref{eq:susceptibility3a}). (e) Calculated values of the mixing coefficients of the exciton states $C^{1s}_{j}(E)$ as a function of the applied electric field.}
    \label{fig:Fig_GaAs_E}
\end{figure}

The experiments presented in this Section were carried out on an GaAs layer with thickness of 10~$\mu$m grown by the gas-phase epitaxy on a semi-insulating GaAs substrate with (001)-orientation~\cite{Pavlov05PRL, Brunne15}. The laser beam was directed along the growth axis of the structure ($\mathbf{k}^{\omega}\parallel[001]$), and the external electric field was applied in the plane of the layer ($\mathbf{E}\perp\mathbf{k}^{\omega}$). In this geometry of the experiment, there is no SHG signal in the absence of an external electric field, but a THG signal is observed at the $1s$-exciton resonance (Figs.~\ref{fig:Fig_GaAs_E}a,b).

For convenience, the equations~(\ref{eq:P3}) and (\ref{eq:P33}) for nonlinear polarization of SHG and THG are  rewritten as follows
\begin{equation}
{P}^{2\omega}_i = \epsilon_0 \chi^{\mathrm{cryst}}_{ijk}(E){E}^{\omega}_j{E}^{\omega}_k + \epsilon_0\chi^{\mathrm{ind}}_{ijkl}(E){E}_j^{\omega}{E}_k^{\omega}{e}_l +
\epsilon_0\chi^{\mathrm{ind}}_{ijklm}(E,k^{\omega}){E}_j^{\omega}{E}_k^{\omega}k_l^{\omega}{e}_m,
\label{eq:SHGind}
\end{equation}
\begin{equation}
{P}^{3\omega}_i = \epsilon_0\chi^{\mathrm{cryst}}_{ijkl}(E){E}_j^{\omega}{E}_k^{\omega}{E}_l^{\omega}+
\epsilon_0\chi^{\mathrm{ind}}_{ijklm}(E){E}_j^{\omega}{E}_k^{\omega}{E}_l^{\omega}{e}_m +
\epsilon_0\chi^{\mathrm{ind}}_{ijklmn}(E,k^{\omega}){E}_j^{\omega}{E}_k^{\omega}{E}_l^{\omega}k_m^{\omega}{e}_n,
\label{eq:THGind}
\end{equation}
where $\mathbf{E} = E\mathbf{e}$ is the applied electric field, $E$ is its amplitude and $\mathbf{e}$ is the unit vector specifying its direction. The electric field can affect differently the nonlinear polarization. First, it can isotropically change the values of components of the nonlinear susceptibility tensor $\chi^{\mathrm{cryst}}(E)$ without changing the symmetry of the system. In this case, the changes introduced by the field into the SHG and THG signals are determined only by the amplitude of the electric field $E$ and do not depend on its direction. Secondly, the electric field $\mathbf{E}$, being a polar vector, can lower the symmetry. These changes are taken into account by introducing tensors of the higher rank $\chi^{\mathrm{ind}}_{ijkl}(E)$, $\chi^{\mathrm{ind}}_{ijklm}(E,k^{\omega})$, $\chi^{\mathrm{ind}}_{ijklm}(E)$ and $\chi^{\mathrm{ind}}_{ijklmn}(E,k^{\omega})$.

As can be seen from Fig.~\ref{fig:Fig_GaAs_E}a, when the electric field is applied, the resonance signal of SHG appears at an energy of 1.516~eV near the $1s$-exciton resonance. The absorption of light is observed at the energy ${\cal E}_{1s} = 1.5152$~eV, the energy shift of the SHG signal is related to the exciton-polariton effect. This effect manifests itself much more strongly in  ZnSe crystal (see Section~\ref{sec:ZnSe_polariton}), in which the exciton has the larger oscillator strength. The integral intensity of SHG grows approximately quadratically with increasing field, see Fig.~\ref{fig:Fig_GaAs_E}c. Unlike SHG, the THG signal is  observed even in the absence of the field, and its intensity in the field decreases, see Figs.~\ref{fig:Fig_GaAs_E}b,d.

The microscopic mechanism of the action of the electric field on the exciton SHG and THG is due to the fact that the electric field mixes the $s$ and $2p$-states with  different symmetry~\cite{Brunne15}. The mixed-state wave function has three contributions $\Psi^{\mathrm{f}} ({\cal E}^{\mathrm{f}}) = C^{\mathrm{f}}_{1s}(E) \Psi_{1s} + C^{\mathrm{f}}_{2s}(E) \Psi_{2s} + C^{\mathrm{f}}_{2p}(E) \Psi_{2p}$, which include unperturbed wave functions $\Psi_j$ with mixing field-dependent coefficients $j = 1s, 2s, 2p$. The magnitudes of these coefficients are shown in Fig.~\ref{fig:Fig_GaAs_E}e. It can be shown for the geometry of the experiment $\mathbf{k}^{\omega} \parallel \mathbf{z}$ and $\mathbf{E} \perp \mathbf{k}^{\omega}$,  i.e.  $\mathbf{E}=E(e_x,e_y,0)$,  that only the induced susceptibility contributes to the SHG signal
\begin{equation}
\chi^{\mathrm{ind}}_{xxxx} = \chi^{\mathrm{ind}}_{yyyy} = 2\chi^{\mathrm{ind}}_{yxyx} = 2\chi^{\mathrm{ind}}_{xxyy},
\end{equation}
\begin{equation}
\chi^{\mathrm{ind}}_{xxxx}({\cal E}^{1s},E) = \chi^{\mathrm{ind}}_{yyyy}({\cal E}^{1s},E) \propto C^{1s}_{2p}(E) C^{1s}_{1s}(E).
\label{eq:susceptibility2a}
\end{equation}
In the case of the THG signal, it is sufficient to take into account only the crystallographic contribution to the nonlinear susceptibility
\begin{equation}
\chi^{\mathrm{cryst}}_{ijkl}({\cal E}^{1s}, E) \propto [C^{1s}_{1s}(E)]^2\approx 1-[C^{1s}_{2p}(E)]^2.
\label{eq:susceptibility3a}
\end{equation}
The decrease in the amplitude of the THG signal is due to the decrease of the $1s$-contribution in the wave function of the exciton state, i.e. with a decrease of $C^{1s}_{1s}(E)$. The model calculations, the results of which are shown by lines in Figs.~\ref{fig:Fig_GaAs_E}c,d, give an excellent agreement with the experiment.

\subsection{Optical second harmonic generation induced by a magnetic field on magneto-excitons in GaAs (orbital quantization)}
\label{sec:magnetic}

To account for the effect of an external magnetic field in the SHG process, the nonlinear polarization ${\mathbf{P}}^{2\omega}$ can be written by analogy to the description of the electric field effect in the form (see the previous Section)
\begin{equation}
{P}^{2\omega}_i = \epsilon_0 \chi^{\mathrm{cryst}}_{ijk}(B){E}^{\omega}_j{E}^{\omega}_k + \epsilon_0\chi^{\mathrm{ind}}_{ijkl}(B){E}_j^{\omega}{E}_k^{\omega}{b}_l +
\epsilon_0\chi^{\mathrm{ind}}_{ijklm}(B,k^{\omega}){E}_j^{\omega}{E}_k^{\omega}{k}_l^{\omega}{b}_m,
\label{eq:SHGindB}
\end{equation}
where $\mathbf{B}=B\mathbf{b}$ is the external magnetic field, $B$ is its amplitude and $\mathbf{b}$ is the unit vector specifying its direction. The need to use the $\mathbf{k}\mathbf{B}$ term describing the nonlinear magneto-optical spatial dispersion, is determined by the details of the microscopic SHG mechanism of  particular excitonic states~\cite{Pavlov05PRL, Saenger06a, Lafrentz13}. The magnetic field breaks the symmetry with respect to the time reversal, which leads to the appearance of new tensors of nonlinear optical susceptibility. The magnetic field acts on electronic states through moving charges, providing an orbital quantization due to the cyclotron motion of electrons and holes. Along with this, the magnetic field acts on the spin states of the electrons and leads to the Zeeman splitting. Both these effects can be manifested in SHG.

\begin{figure}[h!]
    \centering
    \includegraphics[width=10cm]{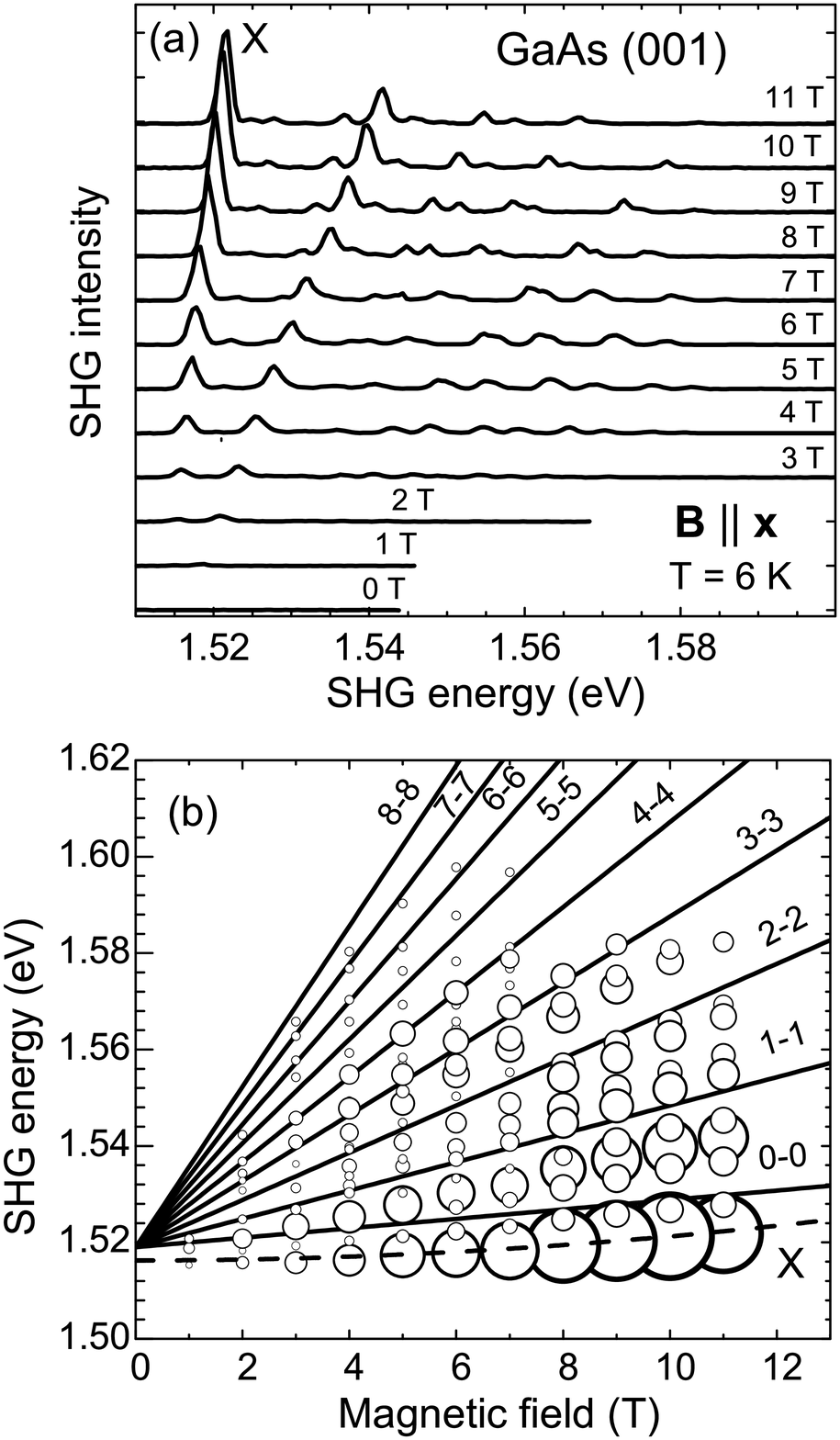}
    \caption{The magnetic-field-induced SHG on exciton states in GaAs. (a) The SHG spectra for various values of the magnetic field in the Voigt geometry $\mathbf{B} \perp \mathbf{k}^{\omega} \parallel [001]$, $\mathbf{E}^{\omega}\parallel \mathbf{E}^{2\omega}$. The spectrum was obtained by scanning the photon energy of laser pulses with 8~ns duration and a repetition rate of 10~Hz. (b) The fan-chart for the magneto-exciton resonances in the SHG spectrum is the dependence of the energy of SHG peaks on the magnetic field. The symbol size illustrates the intensity of the exciton lines. The dashed line shows the literature data for the diamagnetic shift of the $1s$-exciton. The solid lines give the energy of the optical transitions between the Landau levels of free electrons and holes calculated by using Eq.~\eqref{eq:e2} for $N_e$\,=\,$N_h$. The states of the magneto-excitons are shifted below these energies because of the Coulomb interaction.}
    \label{fig:Fig_GaAs_B}
\end{figure}

\begin{figure}[h!]
    \centering
    \includegraphics[width=11cm]{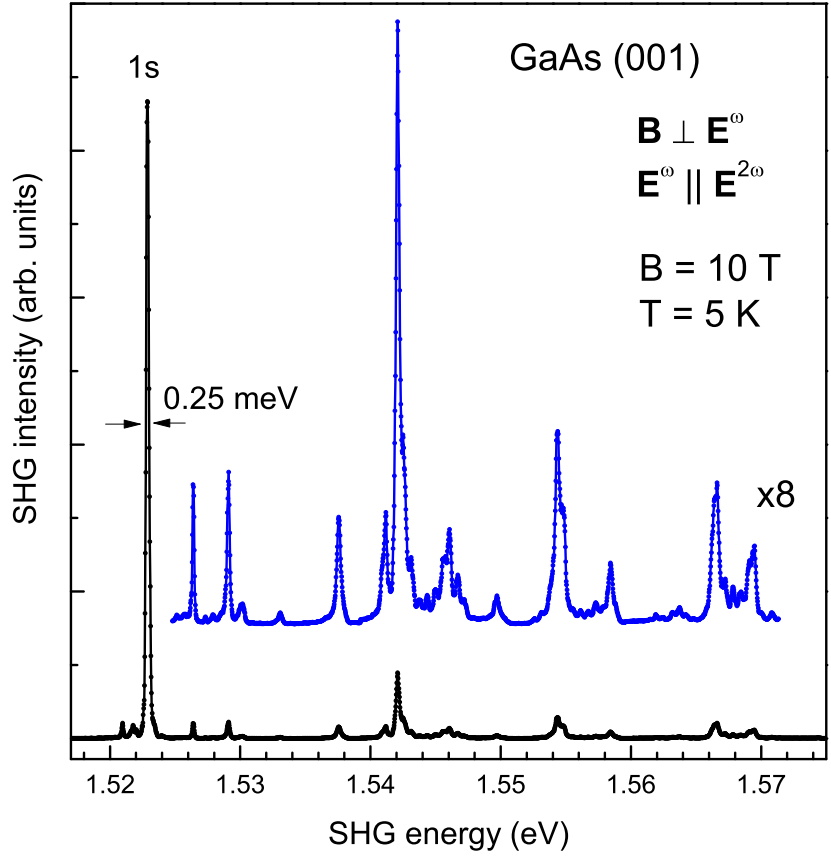}
    \caption{Spectral dependence of the SHG induced by the magnetic field in the Voigt geometry $\mathbf{B} \perp \mathbf{k}^{\omega} \parallel [001]$ on magneto-excitons in GaAs. SHG was excited by spectrally broad laser pulses of 150~fs duration and a repetition rate of 30~kHz. The spectrum was obtained by superposition of six spectra from laser pulses with six different central photon energies.}
    \label{fig:Fig_GaAs_B_fs}
\end{figure}

The SHG induced by a magnetic field on the magneto-exciton states in GaAs is shown in Fig.~\ref{fig:Fig_GaAs_B}a. The SHG signal appears in the field and increases in intensity as a square of the magnetic field. As the field grows, more states appear in the spectrum, which shift toward higher energies, as shown in Fig.~\ref{fig:Fig_GaAs_B}b. A similar behavior is observed in the linear optical spectra of magnetoexcitons~\cite{Seisyan1, Seisyan2}. Undoubtedly, the SHG spectrum should be richer than the linear absorption spectrum, since states relevant to two-photon excitation can appear. As can be seen from Fig.~\ref{fig:Fig_GaAs_B}, the experimental picture is very complicated, and a large number of observed lines requires a deeper analysis. In this case, we limit ourselves to identifying optical transitions between the Landau levels for electrons and holes, described by the expression for the energy of orbital quantization
\begin{equation}
E=E_{g}+\frac{e\hbar }{c}\left[ \frac{1/2+N_{e}}{m_{e}}+\frac{%
1/2+N_{h}}{m_{h}}\right] B,
\label{eq:e2}
\end{equation}
where $E_{g} = 1.519$~eV is the GaAs band gap, $m_{e} = 0.067m_{0}$ and $m_{hh} = 0.51m_{0}$ are the effective electron and heavy-hole masses, $N_{e} = N_{h} = 0, 1, 2, ... $ are the numbers of the Landau levels. Comparing Fig.~\ref{fig:Fig_GaAs_B}a with Fig.~\ref{fig:Fig_GaAs_B_fs} one can see that the measurement with femtosecond pulses and a high spectral resolution makes it possible to detect a much larger number of exciton resonances with very narrow spectral widths not exceeding 0.25~meV.

It is interesting to note that SHG signals induced by the magnetic field on the exciton states in GaAs were measured at lattice temperatures up to 200~K~\cite{Saenger06a}. Qualitatively similar results were obtained on excitons in CdTe and Cd$_{1-x}$Mg$_x$Te ($x=$ 0.01-0.08)~\cite{Saenger06a}.

\subsection{Second harmonic generation induced by a magnetic field on excitons in ZnO (the magneto-Stark effect)}
\label{sec:ZnO}

A detailed experimental and theoretical study of various mechanisms responsible for the magnetic-field-induced SHG on exciton states was carried out for the model and well-studied semiconductor ZnO~\cite{Lafrentz13, Lafrentz13PRL}. ZnO crystal has a noncentrosymmetric crystallographic structure of the wurtzite type (the crystallographic point group $6mm$, the space group $P6_3mc$). The investigated crystal was grown by the hydrothermal method. The sample had the thickness of 500~$\mu$m and the [0001]-orientation, i.e. its optical axis $c$ was perpendicular to the sample plane.

\begin{figure}[h!]
    \centering
    \includegraphics[width=16cm]{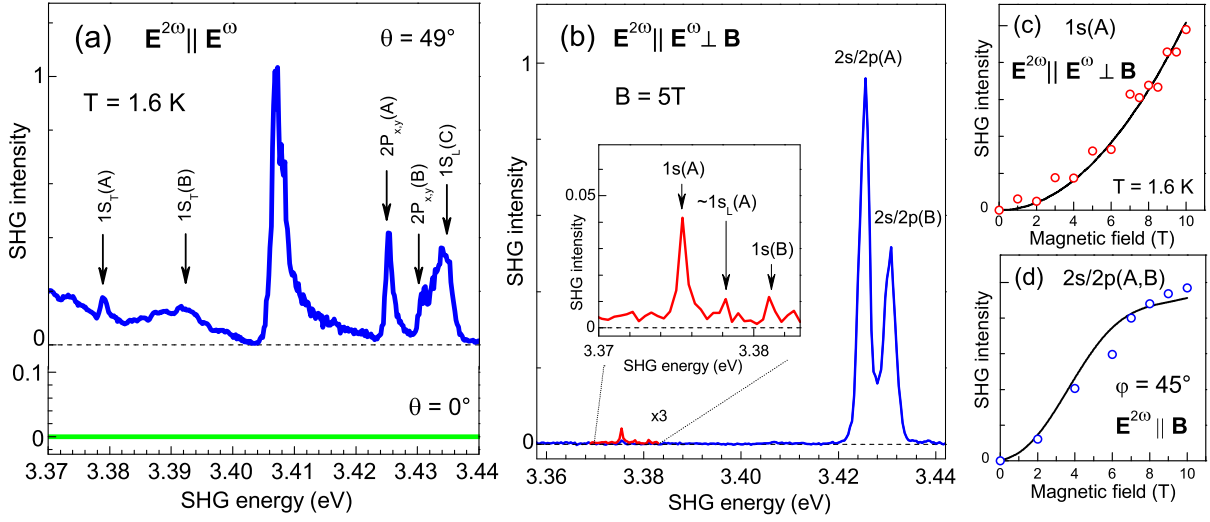}
    \caption{The spectra of SHG signals in ZnO, obtained by scanning the photon energy for laser pulses of 8~ns duration and a repetition frequency of 10~Hz. (a) Spectra of the SHG crystallographic contribution without an external magnetic field. The signal is absent for $\mathbf{k}^{\omega} \parallel [0001]$ and manifests itself at an inclination angle $\theta = 49^\circ$ between the optical axis $c$ and the wave vector $\mathbf{k}^{\omega}$. Literature values of  energies of  A, B, and C-exciton states are shown by arrows. (b) The spectrum of a magnetic-field-induced SHG in the Voigt geometry  $\mathbf{B} \perp \mathbf{k}^{\omega} \parallel [0001]$. (c, d) Dependences of the integrated intensity of SHG signals on the magnetic field for  $1s$(A) and $2s/2p$(A,B)-exciton lines.}
    \label{fig:Fig_ZnO_B}
\end{figure}

The crystallographic contribution to the SHG signal vanishes when the laser beam propagates parallel to the optical axis $c$ ($\theta = 0^\circ$, see Fig.~\ref{fig:Fig_ZnO_B}a), because SHG is symmetry forbidden in the ED approximation. The crystallographic SHG becomes allowed when the sample is rotated by an angle $\theta $ between the $c$ axis and the wave vector $\mathbf{k}^{\omega}$. Figure~\ref{fig:Fig_ZnO_B}a shows a rich spectrum of exciton resonances for $\theta = 49^\circ$.

The SHG signals at the exciton $1s$ and $2s/2p$-states can be induced by a magnetic field in the orientation $\theta = 0^\circ $, as shown in Figs.~\ref{fig:Fig_ZnO_B}b-d. It is surprising that, despite the large oscillator strength of the $1s$-exciton state, the intensity of the relevant SHG signal is almost 100 times weaker than for the $2s/2p$-line (see Fig.~\ref{fig:Fig_ZnO_B}b). Obviously, this should be due to the different effectiveness of the corresponding SHG mechanisms. A detailed experimental and theoretical analysis, carried out in the paper~\cite{Lafrentz13PRL}, showed that a strong signal appears due to the magneto-Stark effect, which mixes the  $2s$ and $2p$ exciton states. The magneto-Stark effect was predicted theoretically in 1955~\cite{Samoilovich} and demonstrated experimentally in 1960 in the linear optical response of excitons in CdS and later in Cu$_2$O~\cite{Thomas0, Thomas, Gross61, Gross68}. The effect arises when the exciton moves at an angle to the magnetic field, and the electron and the hole are acted upon by the Lorentz force in opposite directions. In a coordinate system tied to the center of mass of an exciton, this is equivalent to the appearance of an effective electric field $\mathbf{E}_{\mathrm{eff}}$
\begin{eqnarray}
\mathbf{E}_{\mathrm{eff}}=\frac{\hbar }{M_{\mathrm{exc}}}\left[
\mathbf{k_{\mathrm{exc}}}\times \mathbf{B} \right],
\label{Magneto-stark}
\end{eqnarray}
where $M_{\mathrm{exc}}=m_{e}+m_{h}$ is the exciton effective mass.
For the coherent SHG process, it is required that the involved optical transitions are allowed both for two-photon absorption and for one-photon emission. However, in the ED approximation, the $2s$-exciton state can be addressed only by a one-photon process, whereas the $2p$-exciton state by only a two-photon process. Thus, each of these states alone can not participate in the SHG process. However, their mixing due to the effective electric field $\mathbf{E}_{\mathrm{eff}}$ leads to the formation of a hybrid $2s/2p$-state that proves to be active for the process of second harmonic generation. It should be noted that in the case of the optical harmonics generation, the magneto-Stark effect manifests itself much more pronounced than in linear optics.

\begin{figure}[h!]
    \centering
\includegraphics[width=16cm]{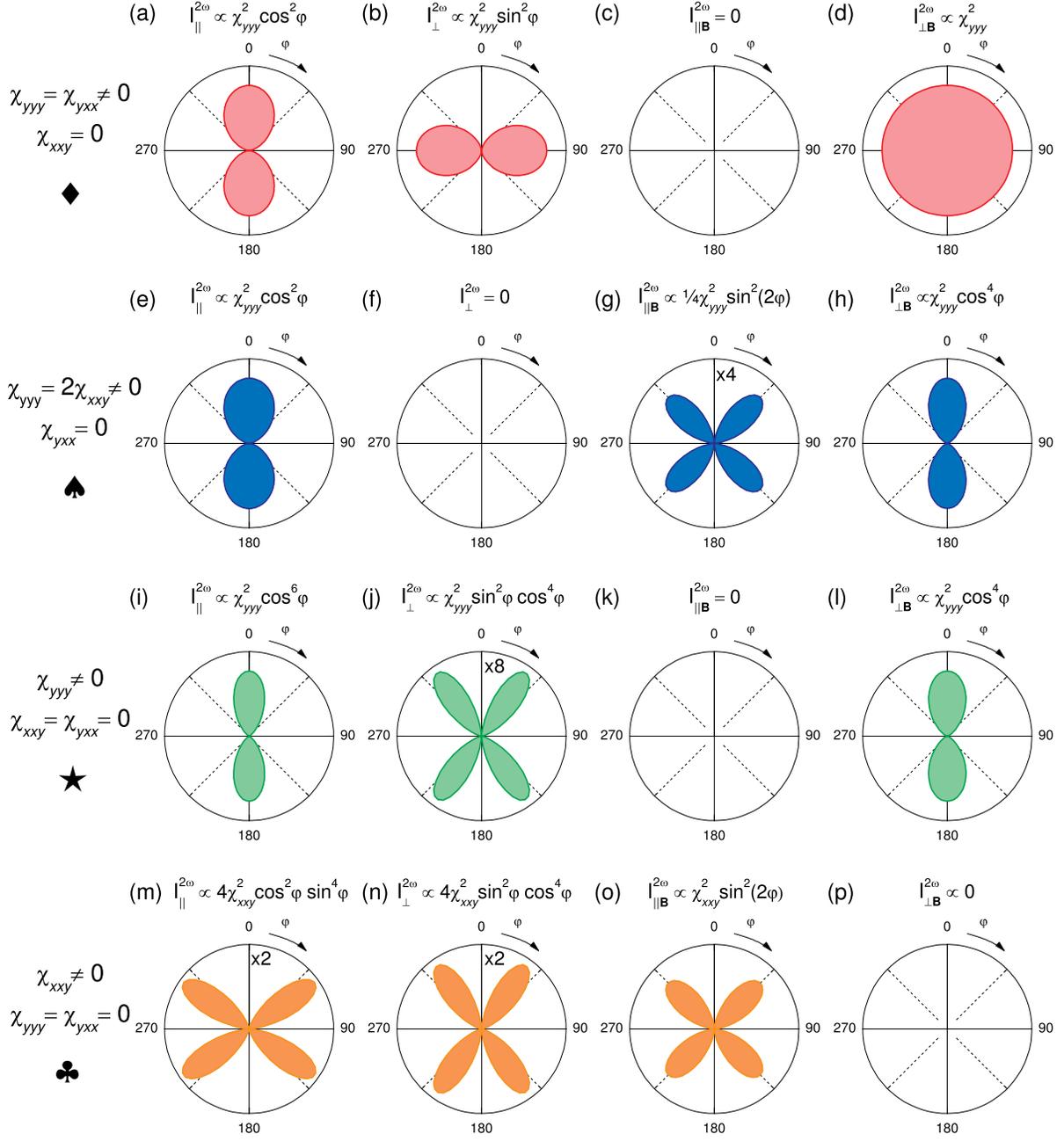}
    \caption{Calculation of rotational anisotropies of SHG in ZnO for different orientations:
$I_\parallel$ for $\mathbf{E}^{\omega}\parallel\mathbf{E}^{2\omega}$,
$I_\perp$\ for $\mathbf{E}^{\omega}\perp\mathbf{E}^{2\omega}$,
$I_{\parallel\mathbf{B}}$ for $\mathbf{E}^{2\omega}\parallel\mathbf{B}$
and $I_{\perp\mathbf{B}}$ for $\mathbf{E}^{2\omega}\perp\mathbf{B}$.
Symbols show the correspondence of diagrams and mechanisms from Table~\ref{tab:Mechanisms}:
(a)-(d) $\blacklozenge$ --- the spin Zeeman effect; (e)-(h) $\spadesuit$ --- the Stark effects, magneto-Stark and orbital Zeeman effects; (i)-(l) $\bigstar$ --- the spin Zeeman effect on the $2p_y$ state; (m)-(p) $\clubsuit$ --- the spin Zeeman effect on the $ 2p_x $ state.
     }
    \label{fig:Fig_ZnO_aniso}
\end{figure}

To identify the various mechanisms for optical second harmonic generation on excitons in ZnO, the rotational anisotropy method described in Section~\ref{sec:experiment} can be used. The rotational anisotropy measurements are usually performed by rotating either parallel or crossed oriented linear polarizers in the excitation and detection channels (see Section~\ref{sec:experiment}); other configurations can be used too. For example, it is possible to rotate only one of the polarizers, while keeping another parallel or perpendicular to the magnetic field. In combination with a detailed microscopic analysis of various mechanisms, rotational anisotropies allow one to reliably identify possible mechanisms, or to separate their contributions in the case of the involvement of several mechanisms at a specific resonance. A detailed consideration of these issues can be found in the paper~\cite{Lafrentz13}, and the results are presented in Table~\ref{tab:Mechanisms}, where several mechanisms and corresponding components of the nonlinear susceptibility tensors are presented. Rotational anisotropies corresponding to these mechanisms are shown in Fig.~\ref{fig:Fig_ZnO_aniso}. In paper~\cite{Lafrentz13}, three possible mechanisms for the SHG induced by the magnetic field are discussed: spin Zeeman effect, Zeeman orbital effect, and magneto-Stark effect. All the considered SHG mechanisms were manifested on exciton states in ZnO. For example, the spin Zeeman effect corresponds to the appearance of the SHG signal on the $1s$-exciton (see Table~\ref{tab:Mechanisms}).

\subsection{Optical second harmonic generation induced by a magnetic field on excitons in (Cd,Mn)Te (spin quantization)}
\label{sec:CdMnTe}

The diluted magnetic semiconductor (Cd,Mn)Te is a good candidate for demonstrating SHG contributions related to the spin splitting of excitons. (Cd,Mn)Te is a structural analogue of the diamagnetic semiconductors CdTe and (Cd,Mg)Te, in which the SHG induced by the magnetic field due to the orbital quantization~\cite{Saenger06a} was observed. The magnetic ions Mn$^{2+}$, which have an uncompensated spin moment $S = 5/2$ on the $3d^5$-electron shell, replace Cd$^{2+}$ ions in the CdTe crystal. Moreover, substitution can take place in any proportion with the formation of a continuous series of solid solutions Cd$_{1-x}$Mn$_{x}$Te, up to MnTe. The strong exchange interaction of electronic states at the edge of the conduction band and the valence band with localized spins of Mn$^{2+}$ ions leads to the giant spin splitting of both the band states and relevant exciton states~\cite{DMS_book, Furdyna}. In the material under study Cd$_{0.84}$Mn$_{0.16}$Te ($E_{g}=1.869$~eV), this splitting reaches 120~meV for $B = 10$~T at $T = 4.5$~K. Phenomenologically, the giant spin splitting is described by a modified Brillouin function$\texttt{B}_{\frac{5}{2}}$:
\begin{equation}
E_{GZ}(S, J)=xS_{0}N_{0}(\frac{\beta}{3} J-\alpha
S)\texttt{B}_{\frac{5}{2}}\left[\frac{5\mu_{B}g_{Mn}B}{2k_{B}(T_{Mn}+T_{0})}\right],
\label{eq:GZS}
\end{equation}
where $g_{Mn}=2$, $k_{B}$ is the Boltzmann constant, and $T_{Mn}$ is the spin temperature of the Mn$^{2+}$ ion system. In our experiment, the spin temperature is equivalent to the crystal lattice temperature $T_{Mn}=T$, $S=1/2$ is the electron spin in the conduction band,  $J=3/2$ is the magnetic moment of holes in the valence band. $S_{0}$ and $T_{0}$ are phenomenological parameters taking into account the antiferromagnetic interaction of Mn-Mn, $N_{0}\alpha=220$~meV and $N_{0}\beta=-880$~meV are exchange integrals for electrons in the conduction band and holes in the valence band interacting with spins of Mn$^{2+}$ ions, respectively.

\begin{figure}[h!]
    \centering
    \includegraphics[width=14cm]{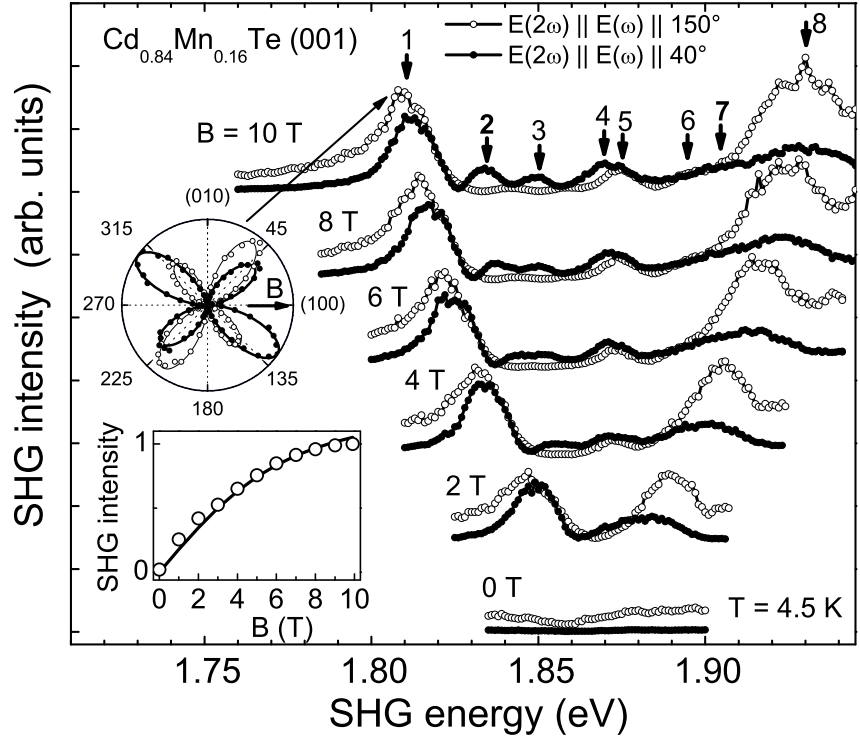}
    \caption{Exciton spectra of SHG in a diluted magnetic semiconductor Cd$_{0.84}$Mn$_{0.16}$Te under conditions of giant spin splitting due to the exchange interaction of excitons with magnetic Mn$^{2+}$ ions. The magnetic field is applied in the Voigt geometry  $\mathbf{B} \perp \mathbf{k}^{\omega} \parallel [001]$. The spectra were obtained by scanning the energy of photons for laser pulses of 8~ns duration and a repetition rate of 10~Hz. Inset shows the dependence of the integrated SHG intensity on the magnetic field. This dependence reflects the increase in magnetization in accordance with the modified Brillouin function according to Eq.~\eqref{eq:GZS}.
The presented rotational anisotropies of the SHG signals are measured in the magnetic field of 10~T at an energy of 1.81~eV for two polarizer orientations: $\mathbf{E}^{\omega} \parallel \mathbf{E}^{2\omega}$ is shown by open circles, and $\mathbf{E}^{\omega} \perp \mathbf{E}^{2\omega}$ is shown by closed circles.
     }
    \label{fig:Fig_CdMnTe_A}
\end{figure}

\begin{figure}[h!]
    \centering
    \includegraphics[width=14cm]{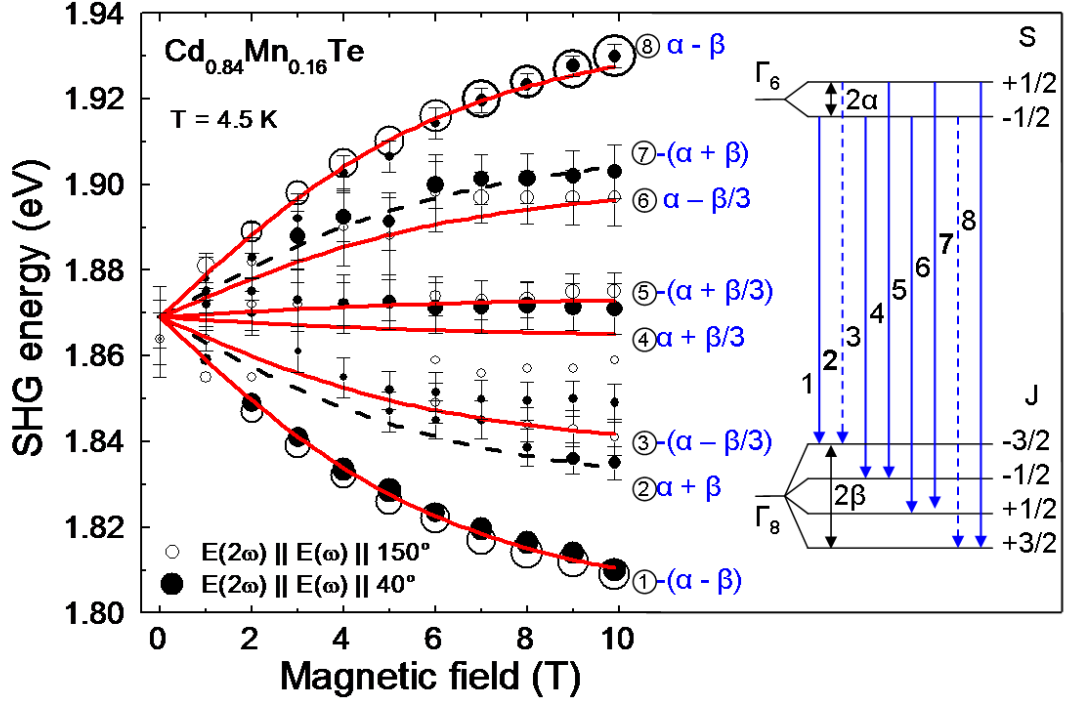}
    \caption{Dependence of the positions of SHG exciton lines in Cd$_{0.84}$Mn$_{0.16}$Te, presented in Fig.~\ref{fig:Fig_CdMnTe_A}, on the magnetic field (symbols). The lines show the calculated energies of the optical transitions shown in inset. The calculation was carried out using Eq.~\eqref{eq:GZS} with the parameters $S_{0}=0.8$ and $T_{0}=6.5$~K. In inset solid arrows show single-photon transitions observed both in linear optics and in nonlinear SHG spectroscopy. The dashed arrows indicate two-photon transitions observed only in the SHG spectra.}
    \label{fig:Fig_CdMnTe_B}
\end{figure}

A set of SHG spectra induced by an external magnetic field in Cd$_{0.84}$Mn$_{0.16}$Te is shown in Fig.~\ref{fig:Fig_CdMnTe_A}~\cite{Saenger06PRL,Saenger06b}. In the chosen geometry of the experiment $\mathbf{k}^{\omega} \parallel [001]$, the SHG signal vanishes in the zero magnetic field. As the magnetic field grows, SHG signals appear on the exciton states. In the field $B = 10$~T, eight lines can be identified, which correspond to optical transitions between the different spin states of electrons and holes shown in inset of Fig.~\ref{fig:Fig_CdMnTe_B}. It is interesting to note that transitions 2 and 7 require changes in the angular momentum by $\pm 2$ and, therefore, are not observed in single-photon processes, but they are manifested in the SHG spectra. In contrast to diamagnetic semiconductors, where the increase in SHG intensity is proportional to the square of the magnetic field, in (Cd,Mn)Te this growth is proportional to the magnetization of the system of magnetic ions. This is illustrated by inset in Figure~\ref{fig:Fig_CdMnTe_A}. The spectral shifts of the lines in the magnetic field, shown in Fig.~\ref{fig:Fig_CdMnTe_B}, are well described by the modified Brillouin function~\eqref{eq:GZS}. This result shows that in the diluted magnetic semiconductor Cd$_{0.84}$Mn$_{0.16}$Te, the spin splitting induced by the magnetic field is the dominant mechanism of the observed optical second harmonic generation.

The magnetic-field-induced SHG can be used for optical measurement and identification of magnetic ordering. Such studies were carried out on magnetic semiconductors EuTe and EuSe~\cite{Kaminski09, Kaminski10, Lafrentz10, Lafrentz12}. These materials have a simple cubic lattice with a center of inversion, in which SHG is forbidden in the ED approximation. Nevertheless, it turned out that SHG is possible due to MD mechanisms in combination with a magnetization induced by an external magnetic field.

\subsection{Optical second harmonic generation on exciton-polariton in ZnSe}
\label{sec:ZnSe_polariton}

\begin{figure}[h!]
    \centering
    \includegraphics[width=14cm]{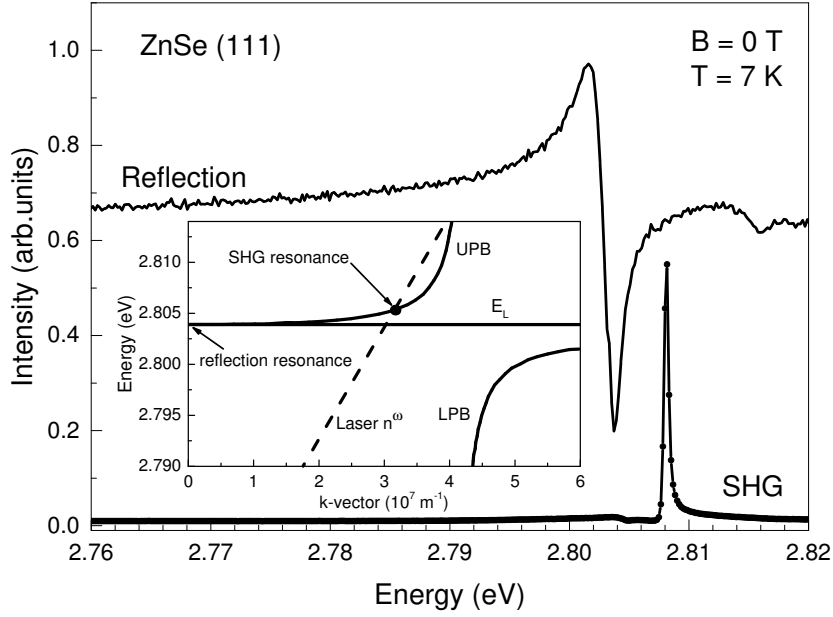}
    \caption{SHG and light reflection spectra on the exciton-polariton in ZnSe. In the experimental geometry  $\mathbf{k}^{\omega} \parallel [111]$, the SHG crystallographic contribution is allowed in the ED approximation. Inset shows schematically the spectral shift between resonances in reflection and SHG. The intersection point of the dispersion dependence of the laser pump with the upper polariton branch (UPB) is noted by the dot. LPB is the lower polariton branch; $E_L$ is the dispersion of the longitudinal branch. SHG was excited by spectrally broad laser pulses of 150~fs duration and a repetition rate of 30~kHz with a maximum at 1.404~eV.}
    \label{fig:Fig_ZnSe_SHG_polariton}
\end{figure}

In Figure~\ref{fig:Fig_ZnSe_SHG_polariton}, the SHG spectrum on the exciton in ZnSe is compared with the reflection spectrum. To measure the reflection, a spectrally broadband halogen lamp source was used. In the reflection spectrum, a strong exciton resonance with a minimum for the photon energy of 2.8037~eV is seen. The narrow resonance in the SHG spectrum is shifted by 4.5~meV toward higher energies and has a maximum at 2.8082~eV. This shift is related to the features of the exciton-polariton dispersion. As shown schematically by inset in Fig.~\ref{fig:Fig_ZnSe_SHG_polariton}, the phase matching conditions for SHG are satisfied at the point where the $n^{\omega}$ dispersion of the laser light intersects with the upper polariton branch (UPB). A resonant enhancement of SHG occurs at this energy. Because of the exciton-polariton dispersion, this point is shifted in energy from the minimum in the spectrum of linear light reflection. This shift of 4.5~meV manifested itself very clearly in ZnSe because of the large exciton oscillator strength (the exciton binding energy is 20~meV). As noted in Section~\ref{sec:electric}, an analogous shift was observed in GaAs, but it was only 0.8~meV because of the smaller exciton binding energy of 4.2~meV. We note that experiments on the measurement of SHG and THG, as well as experiments on two- and three-photon light absorption~\cite{Froehlich94}, may become powerful supplemental tools for studying the exciton-polariton dispersion and its modification in external electric and magnetic fields.

\subsection{Giant growth of the optical third harmonic intensity on the $1s$-exciton in GaAs in a magnetic field}
\label{sec:THG_GaAs}

To take into account the effect of the external magnetic field $\mathbf{B}$ on the THG process, the nonlinear polarization ${\mathbf{P}}^{3\omega}$ can be written as:
\begin{equation}
{P}^{3\omega}_i = \epsilon_0 \chi^{\mathrm{cryst}}_{ijkl}(B){E}^{\omega}_j{E}^{\omega}_k{E}^{\omega}_l + \epsilon_0\chi^{\mathrm{ind}}_{ijklm}(B){E}_j^{\omega}{E}_k^{\omega}{E}^{\omega}_l{b}_m +
\epsilon_0\chi^{\mathrm{ind}}_{ijklmn}(B,k^{\omega}){E}_j^{\omega}{E}_k^{\omega}{E}^{\omega}_l{k}_m^{\omega}{b}_n.
\label{eq:THGindB}
\end{equation}

\begin{figure}[h!]
    \centering
    \includegraphics[width=12cm]{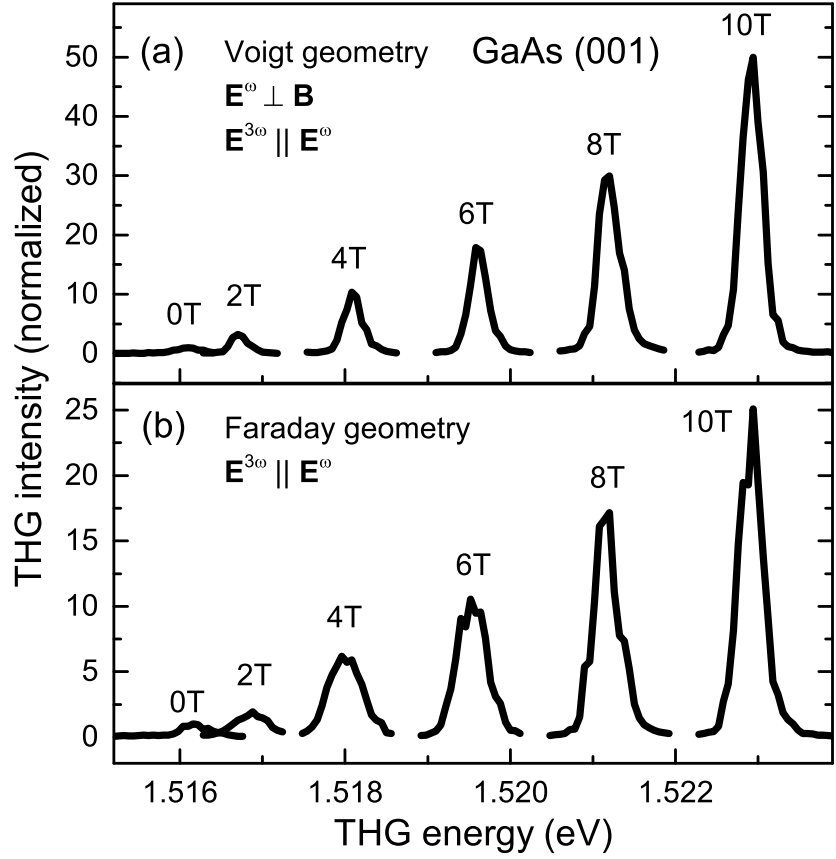}
    \caption{Spectra of THG signals in GaAs at $1s$-exciton resonance as a function of the magnetic field in the Voigt geometry $\mathbf{B} \perp \mathbf{k}^{\omega} \parallel [001]$ (a), and Faraday geometry $\mathbf{B} \parallel \mathbf{k}^{\omega} \parallel [001]$ (b). The spectra were obtained by scanning the energy of photons for laser pulses of 8~ns duration and a repetition rate of 10~Hz. The spectra are normalized to the peak intensity of the THG at $B = 0$, so the numerical values on the vertical scale correspond to the factor of the THG gain in the applied magnetic field with respect to $B = 0$.}
    \label{fig:Fig_GaAs_THG_B}
\end{figure}

Above in this paper, when studying the processes of second harmonic generation in the magnetic field, we choose the experimental geometry in which the SHG crystallographic contribution is symmetry forbidden. As a result, the intensity of SHG increases with increasing field, starting from zero values at $B = 0$. In the case of the optical third harmonic generation, the situation is different, since this process is symmetry allowed in all materials and in all orientations. Induced contributions from the external field $\chi^{\mathrm{ind}}_{ijklm}(B)$ and $\chi^{\mathrm{ind}}_{ijklmn}(B,k^{\omega})$ can interfere with the THG crystallographic contribution $\chi^{\mathrm{cryst}}_{ijkl}$. They can often be ignored, but one has to take into account a dependence of the crystallographic contribution itself on the magnetic field $\chi^{\mathrm{cryst}}_{ijkl}(B)$.
The results of this experimental situation are shown in Fig.~\ref{fig:Fig_GaAs_THG_B}. THG was measured in GaAs for two orientations of the external magnetic field. In both cases, there was a very strong increase in the THG intensity, which in the field of 10~T reaches a factor of~50 for the Voigt geometry and of~25 for the Faraday geometry. The performed model analysis showed that these effects for the THG enhancement can be explained by an anomalous dispersion in the region of exciton-polariton states, an increase in the exciton oscillator strength in the magnetic field, and the reabsorption of the THG signal upon propagation in a GaAs sample. Model estimates explain the THG enhancement by a factor of tens and identify the growth of the oscillator strength as the main amplification mechanism. A qualitative confirmation of the adequacy of this model is the fact that the THG enhancement is much weaker in CdTe and is practically absent in ZnSe because in these materials the magnetic fields up to 10~T only weakly increase the oscillators strength of their excitons.

\subsection{Optical second harmonic generation on quantum-confined excitons in ZnSe/BeTe quantum wells}
\label{sec:QW}

\begin{figure}[h!]
    \centering
    \includegraphics[width=12cm]{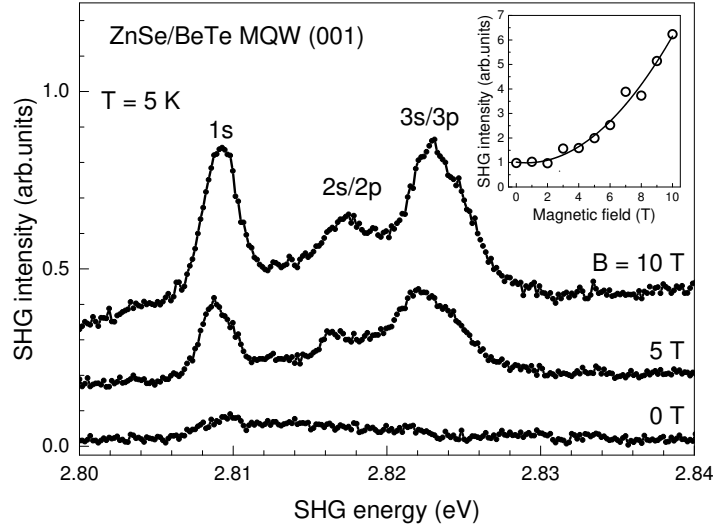}
    \caption{SHG spectra for the quantum well structures ZnSe/BeTe (20~nm/10~nm, 10 periods) in magnetic fields 0, 5~T and 10~T in the Voigt geometry  $\mathbf{B} \perp \mathbf{k}^{\omega} \parallel [001]$, $\mathbf{E}^{\omega} \parallel \mathbf{E}^{2\omega}$.  SHG was excited by spectrally broad laser pulses of 150~fs duration and a repetition frequency of 30~kHz with a maximum at the photon energy of 1.410~eV. Inset shows the magnetic field dependence of the SHG signal integrated in the spectral range 2.805-2.830~eV; the line shows the fitting curve for the quadratic dependence of the SHG intensity on the magnetic field, $I^{2\omega} \propto B^2$.}
    \label{fig:Fig_ZnSe_BeTe_MQW}
\end{figure}

Exciton spectroscopy by the SHG method can also be extended to the study of excitons in low-dimensional semiconductor heterostructures, but up to now no such studies were reported. Our first results for the ZnSe/BeTe structure with quantum wells (20~nm/10~nm, 10 periods) grown by molecular-beam epitaxy on the (001)-oriented GaAs substrate are presented in Fig.~\ref{fig:Fig_ZnSe_BeTe_MQW}. This heterostructure has a band structure of type II, in which the electrons are  quantized in the conduction band of the ZnSe layers, and the holes in the valence band of the BeTe layers~\cite{Platonov99}. In this case, electrons due to the Coulomb interaction create a metastable state for the holes in the ZnSe layer. In sufficiently wide quantum wells (> 15~nm), this allows observing in the reflection and photoluminescence spectra of quasi-two-dimensional excitons formed by such carriers~\cite{Platonov98, Zaitsev97, Maksimov99}. The energy $E_g=2.81$~eV of these excitons is near the band gap of ZnSe.

In the experiment shown in Fig.~\ref{fig:Fig_ZnSe_BeTe_MQW}, the wave vector of the exciting light is directed parallel to the growth axis of the structure $\mathbf{k}^{\omega} \parallel [001]$. As already discussed above, under these conditions SHG is forbidden in the ED approximation, which agrees with the presence of only very weak SHG signals in the absence of an external magnetic field. The presence of these weak signals can be related to a certain decrease in the symmetry of the system due to an electric field induced by spatially separated electrons and holes and directed along the growth axis of the structure. A significant increase of the SHG intensity, which is quadratic in the magnetic field (see inset in Fig.~\ref{fig:Fig_ZnSe_BeTe_MQW}), is observed when the field is applied in the Voigt geometry. The identification of $1s$, $2s/2p$ and $3s/3p$ exciton states is based on their diamagnetic shift. Qualitatively, the magnetically induced SHG in a ZnSe/BeTe quantum-well structure is analogous to that in bulk diamagnetic semiconductors GaAs, CdTe, and ZnO~\cite{Saenger06a, Lafrentz13}. A detailed analysis of different contributions of specific microscopic mechanisms requires further investigation of SHG on quantum-confined excitons in quantum wells. As far as we know, this is the first observation of SHG induced by an external magnetic field on quasi-two-dimensional excitons in semiconductor heterostructures with quantum wells. Due to a such experimental approach, it became possible to study SHG using spectrally broad laser pulses of 150~fs duration and using high spectral resolution in the signal recording channel.
This method opens up wide opportunities for investigating the nonlinear optical properties in low-dimensional heterostructures.

\subsection{Optical second harmonic generation on the Gross exciton in centrosymmetric Cu$_2$O}
\label{sec:Cu2O}

\begin{figure}[h!]
    \centering
    \includegraphics[width=14 cm]{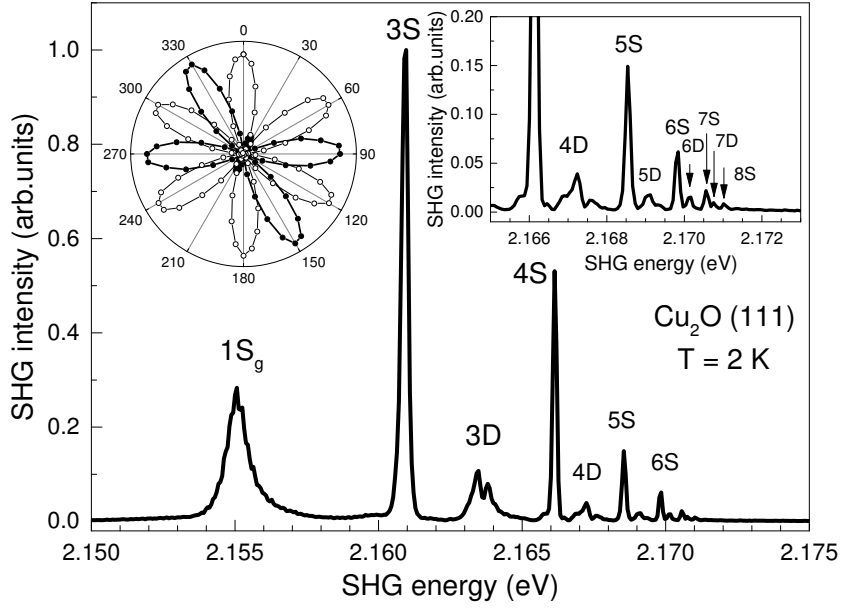}
    \caption{SHG spectrum for Cu$_2$O(111), measured upon excitation by spectrally broad laser pulses of 150~fs duration and a repetition frequency of 30~kHz with a maximum for the photon energy of 1.087~eV, $\mathbf{k}^{\omega} \parallel [111]$, $\mathbf{E}^{2\omega} \parallel \mathbf{E}^{\omega} \parallel [11\overline{2}]$. The ground state of the exciton of the $1S_g$ green series and the excited states of the yellow exciton series are shown. Inset shows rotational anisotropies of SHG measured for the $3S$ line in the geometries $\mathbf{E}^{\omega} \parallel \mathbf{E}^{2\omega}$ (closed circles) and $\mathbf{E}^{\omega} \perp \mathbf{E}^{2\omega}$ (open circles).
}
    \label{fig:Fig_Cu2O}
\end{figure}

Surprisingly, the Cu$_2$O crystal in which the hydrogen-like series of excitations was first observed and became a proof of the existence of excitons~\cite{Gross52}, turned out to be absolutely unique in its properties for exciton spectroscopy. Recently, exciton excited states up to $n = 25$ were found in Cu$_2$O linear absorption spectra~\cite{Kazimierczuk14}. Figure~\ref{fig:Fig_Cu2O} shows the SHG spectrum measured in the (111)-oriented Cu$_2$O using spectrally broad pulses of 150~fs duration. In these experiments, the spectral maximum of the laser pulses was fixed at an energy of 1.087~eV, and the integration time of the spectrum did not exceed 5 min. The narrow lines corresponding to the $1S_g$ state of the green exciton series and the excited states of the excitons of the yellow series up to $8S$ are clearly seen in the spectrum. The observation of a strong SHG signal is very surprising, since the optical second harmonic is symmetry forbidden in the ED approximation for the centrosymmetric crystallographic structure of Cu$_2$O. An analysis of the rotational anisotropies for the SHG signals allows us to conclude that the  EQ process is responsible for the one-photon emission of SHG.

\section{Conclusion}

In this paper, we presented a brief review to show the wide possibilities of studying excitons in semiconductors by the method of optical harmonics generation. This approach allows one to investigate exciton properties that are not accessible by the methods of linear optical spectroscopy, and to provide new information on the energy and spin structure of exciton states. Up to now, these studies have been performed on a limited number of model bulk semiconductors. One can therefore assume that this research  field  is only at the very beginning. There is no doubt that the extension of the research using existing and developing methods of optical harmonics spectroscopy on other semiconductors and nanostructures will bring many new and unexpected results.

\section{Acknowledgments}

This review includes the results of studies performed at the TU Dortmund University (Dortmund, Germany) and the Ioffe Institute of the Russian Academy of Sciences (St. Petersburg, Russia) in the period from 2004 to the present time. We are grateful to colleagues who made a significant contribution to these studies: I. S\"anger, B. Kaminski, A. M. Kalashnikova, M. Lafrentz, V. A. Lukoshkin, A. B. Henriques, D. Brunne, D. Fr\"{o}hlich, D. Feng and A. Farenbruch. We are also grateful to E. L. Ivchenko, M. M. Glazov and M. A. Semina for fruitful discussions.

\newpage
\section*{Figures and Table}

\newpage

\newpage

\newpage

\newpage

\newpage

\newpage

\newpage

\newpage

\newpage

\newpage

\newpage

\newpage

\newpage
\begin{table*}\footnotesize
\begin{tabular}{l|c|c|c|c|c}
Mechanisms & $1s$, $2s$  & $2s/2p_y$  & $2p_z/2p_y$  &   $2p_y$ & $2p_x$ \\
\hline \hline
Stark effect & & $\chi_{yyy}= 2\chi_{xxy}\ne 0$,& & & \\
$D_i^{2\omega}D_j^\omega D_l^\omega$& & & & & \\
$E_y \neq 0$, $B_x = 0$ & & $\chi_{yxx}=0$ $\spadesuit$ & & & \\
\hline
Magneto-Stark effect & & $\chi_{yyy}= 2\chi_{xxy}\ne 0$,& & & \\
$D_i^{2\omega}D_j^{\omega}D_l^{\omega}$& & & & & \\
$E_y = 0$, $B_x \neq 0$ & & $\chi_{yxx}=0$  $\spadesuit$ & & &\\
\hline
Spin Zeeman effect & $\chi_{yyy}= \chi_{yxx}\ne 0$, & & & &  \\
$D_i^{2\omega}D_j^{\omega}Q_l^{\omega,m}$& & & & & \\
$E_y = 0$, $B_x \neq 0$ & $\chi_{xxy}=0$ $\blacklozenge$ & & & & \\
\hline
Spin Zeeman effect & & & &   $\chi_{yyy} \ne 0$, $\bigstar$ & $\chi_{xxy} \ne 0$, $\clubsuit$ \\
$Q_i^{2\omega,m}D_j^{\omega}D_l^{\omega}$& & & & & \\
$E_y = 0$, $B_x \neq 0$ & & & &   $\chi_{xxy}= \chi_{yxx}=0$ & $\chi_{yyy}= \chi_{yxx}=0$\\
\hline
Orbital Zeeman  & & & $\chi_{yyy}= 2\chi_{xxy}\ne 0$, & &\\
effect & & &  & &\\
$Q_i^{2\omega,m}D_j^{\omega}D_l^{\omega}$& & & & & \\
$E_y = 0$, $B_x \neq 0$ & & & $\chi_{yxx}=0$ $\spadesuit$  & &
\end{tabular}
\caption{The SHG mechanisms in applied electric and magnetic fields on $1s(A,B,C)$, $2s(A,B,C)$ and $2p(A,B)$-exciton resonances in hexagonal ZnO. The geometry of the experiment:  $\mathbf{k} \parallel z$, $\mathbf{E}=(0,E_y,0)$, and $\mathbf{B}=(B_x,0,0)$.
Because of the permutation symmetry, $\chi_{xxy} = \chi_{xyx}$. $D$ and $Q$ denote the processes allowed in the ED and EQ approximations, respectively. The symbols correspond to different mechanisms: $\blacklozenge $  to the spin Zeeman effect, $\spadesuit$  to the Stark effects, magneto-Stark and orbital Zeeman effects, $\bigstar$ to the spin Zeeman effect on the $2p_y$ state, and $\clubsuit $ to the spin Zeeman effect on the $2p_x$ state. Rotational anisotropies of the SHG signals corresponding to the SHG in ZnO are shown in Fig.~\ref{fig:Fig_ZnO_aniso}.}
\label{tab:Mechanisms}
\end{table*}

\end{document}